\documentclass{iopjournal}

\usepackage{natbib}
\usepackage{array}
\usepackage{comment}

\begin{document}

\articletype{Paper} 

\title{Making Sense of Quantum Technology: Nuance and Comparison in Non-Expert Stakeholder Communication}

\author{ Floris Löffler$^{1,2,\#}$\orcid{0009-0007-5560-8717}, Lisanne van Veenen$^{1,2,\#}$\orcid{0009-0005-8612-4059}, Muhammad Unggul Karami$^{1,2,*}$\orcid{0009-0000-7820-4335}, and Julia Cramer$^{1,2}$\orcid{0000-0002-4756-0043}}

\affil{$^\#$These authors contribute equally.}

\affil{$^1$Leiden Institute of Physics, Faculty of Science, Leiden University, Leiden, The Netherlands}

\affil{$^2$Science Communication and Society, Leiden University, Leiden, The Netherlands}

\affil{$^*$Author to whom any correspondence should be addressed.}

\email{ukarami@physics.leidenuniv.nl}


\begin{abstract}
The development of quantum technology (QT) is partly driven by a growing group of non-expert stakeholders, such as policymakers, journalists, and philosophers. Although many of these stakeholders have minimal formal training in quantum science, their communications may greatly influence the QT ecosystem, yet research into it is scarce. We therefore explored the use of framing, metaphors, and jargon by non-expert QT stakeholders through a qualitative content analysis of 12 semi-structured interviews. We found that the stakeholders applied a nuanced lens to QT, typically through \textit{socioeconomic} framing. Several types of framing previously unidentified in the literature were found, such as explaining QT by comparison with current (emerging) technologies, for example, artificial Intelligence, in terms of capabilities and implementation in society. The use of stakeholder desiderata in such research was also found to be subsidiary to the stakeholder's occupation. This explorative study is meant to serve as a starting point for future studies of this stakeholder group.
\end{abstract}

\section{Introduction}
Humanity is moving towards the large-scale implementation of quantum technology (QT), with several applications of quantum mechanics rapidly maturing \citep{Grigoryan2025, NaturePortfolio2025, Ramya2025}. The QT ecosystem has experienced important developments in areas such as quantum communications, quantum computing, and quantum sensing \citep{Kimble2008, Coccia2024, Umbrello2024}.  The recent rise in quantum developments, accompanied by large investments, is known as the \textit{second quantum revolution} \citep{Vermaas2019}. While QT could offer society various potential benefits such as new drug discoveries and highly secure communication \citep{Aquina2024-ez}, there are a plethora of risks involved. An outstanding example is the advent of quantum computers, which could endanger cybersecurity by potentially breaking traditional encryption \citep{Castelvecchi2026}. Other risks involve increasing inequality between nations depending on access to these new quantum technologies \citep{TenHolter2022}, and enabling criminal organisations through impenetrable communications \citep{Vermaas2019}. 

Despite its large-scale development, the actual implementation of QT is still in an early, uncertain stage \citep{Coccia2024}. As a result, experts stress that we are at a critical moment for the responsible societal embedding of these technologies \citep{Vermaas2017, TenHolter2022, Purohit2023}. Based on historical analysis, \citet{DeJong2022} formulated a strategy to ``prepare society for the quantum age" (p. 1): A core part of this approach is engaging relevant stakeholders and managing ``unrealistic expectations". 

In this study, we focus on a specific part of the "quantum ecosystem": non-expert stakeholders. In terms of knowledge and engagement, this group is positioned between lay members of the general public and QT experts. Non-expert stakeholders carry their own backgrounds and interests, contributing to how QT development and implementation are shaped \citep{Vermaas2017, Umbrello2024}. Additionally, given their lack of academic background in QT, the interaction and consequent understanding and communication of QT in this stakeholder group could provide starting points for a broader understanding of the general public within this context.

Any technical communication among these stakeholders relies on effective understanding and explanation of abstract concepts such as entanglement or superposition. The language used to explain such technologies can influence stakeholder understanding and might lead to miscommunications \citep{Langer2021terminology}. An important aspect of QT stakeholder language that warrants thorough examination is the use of metaphors, frames and narratives. These linguistic elements often contribute to hype and shape the public understanding of QT \citep{Meinsma2023, Meinsma2025coin, Schyfter2015}. Although research into hype and popularised language use in communication by experts about quantum science and technology is already present \citep{Meinsma2023, karami_quantum_2026}, research is still sparse on non-expert stakeholders. \citet{Meinsma2023} concluded that the language use of non-expert stakeholders deserves separate analysis, since they seem to approach the framing of QT differently from expert stakeholders. Non-experts were found to focus less on certain common frames and to explain underlying quantum concepts to a lesser extent than experts.

We aim to provide a comprehensive, exploratory analysis of the way non-expert stakeholders understand and explain quantum. Increased understanding of this topic can enable non-expert stakeholders to participate in QT \citep{TenHolter2022} and serve their interests at a time when it is crucial to include them \citep{Vermaas2017, Seskir2023}. In this study, we focus on non-expert stakeholders, specifically policymakers, funding agencies, and the media \citep{Umbrello2024}. The next section provides an overview of our theoretical framework and relevant literature. 

\section{Theory}

\subsection{Non-expert stakeholders}\label{subsection Non-expert stakeholders}
Including and engaging stakeholders in the development of new technologies and innovations has seen widespread attention in recent literature, especially concerning the ethical development of quantum technology \citep{Ntorina,Saurabh2025,ArbelaezOssa2024, Barresi2025}. In an exploration of the landscape of quantum stakeholders, \cite{Umbrello2024} defined a stakeholder as ``any individual, group, or entity that has a direct or indirect interest in, is affected by, or can influence the design, development and use of technology, considering its ethical, social and environmental implications (p.13)." In an attempt to categorise the TEDx speakers, \citet{Meinsma2023} defined quantum experts as ``scientists" and ``leaders" in a ``university, institute, research initiative, start-up, or another organization working in`` quantum science and adjacent field (Appendix II p. 1). From that definition, we can infer that stakeholders who are not included are non-expert stakeholders.

\citet{Umbrello2024} identified quantum technology stakeholders and their relation to QT from existing literature, using a Value Sensitive Design (VSD) approach. Based on the working definition above, we consider two of these groups to be expert groups, namely researchers \& developers and educators \& training providers, because their function and descriptions show in-depth knowledge of quantum concepts. The non-expert groups that were identified are end-users; regulators and policymakers; investors and funders; industry partners and suppliers; ethics/societal experts; media and journalists; and the general public \citet{Umbrello2024}. In the context of artificial intelligence (AI), \citet{Langer2021xai} defined the term desiderata: the conglomerate of “stakeholders’ interests, goals, expectations, needs and demands regarding artificial systems”. While this term is not explicitly mentioned in \citep{Umbrello2024}, their study serves as a starting point to discern some of the stakeholders' desiderata.

\subsection{The language of QT stakeholders}\label{The language of QT stakeholders}

To investigate the language use of these QT stakeholders, we used framing, metaphor usage, and jargon as our analytical lenses.

\subsubsection{Framing}\label{framing}

To understand the attitudes people hold when discussing QT, concepts such as ``framing" and ``narratives" can be applied.  These terms are used and defined differently throughout the literature and sometimes overlap \citep{Meinsma2024frames, Pohlmann2024, Bareis2022, Droog2020, Grinbaum2017}. For simplicity, we will solely use the term ``framing" in our content analysis as an umbrella for all related instances in contemporary literature. A useful definition of framing can be found in the work of \citet{Meinsma2024frames}, who applied the following definition from \citet{Entman1993}: “[selection of] some aspects of a perceived reality and making them more salient in a communicating context, in such a way to promote a particular problem definition, causal interpretation, moral evaluation, and/or treatment recommendation (p. 52)”. 

There already exists a wide array of known frames relating to quantum science and technology (QST). In terms of framing, the work of \citet{Meinsma2023} forms the basis of our theoretical framework. Their study entails a content analysis of 501 TEDx talks discussing quantum science or technology to determine the prevalence of previously identified problematic framing of QST. Aside from this work, we took inspiration from the narratives around QT identified by \citet{Pohlmann2024} and \citet{Suter2024} through a content analysis of policy documents, media articles and business reports. They categorised the narratives determined in their study into five main themes: (i) technical aspects \& applications, (ii) politics \& global conflicts, (iii) people \& society, (iv) national technology strategies, and (v) business \& market development. Almost all (sub)narratives within these themes fall into one of the framing categories by \citet{Meinsma2023}. Below, the frames defined by \citet{Meinsma2023} will be described. The discussion is supplemented with relevant narratives from \citet{Pohlmann2024} and other sources. 

\textit{Economic competition versus the public good}. The first theme in the analysis of TEDx talks was the tension between the frame of \textit{economic competition/development} and \textit{social progress}. The ideas of a ``quantum race" or ``international competition" \citep{Pohlmann2024, Suter2024} fall into the frame of \textit{economic competition/development}, while framing in terms of \textit{social progress }entails a broader discussion of how QT can serve the public good. \textit{Social progress} framing involves any subject related to the ``responsible embedding of QT" \citep{Pohlmann2024, Suter2024}. Examples include (in)equality in access to QT and solving problems related to healthcare, climate change, or energy production. Whether \textit{economic competition/development} or \textit{social progress} dominates the other seems to depend on context. \citet{Pohlmann2024} shows an increase in media interest in global tech competition, while interest in the narrative ``People, Society and Culture" was low among all types of communications. In contrast, the analysis of TEDx talks showed a stronger emphasis on \textit{social progress} \citep{Meinsma2023}. 

\textit{Benefits, risks and balanced views of QT.} The second theme the analysis brought forth is the tendency of TEDx speakers to emphasise the benefits over the risks of QT. Additionally, only 4\% of TEDx speakers gave a balanced view of the risks and benefits of QT. Benefits of QT may include secure communication over the quantum internet and drug discovery through simulations performed by quantum computers. Risks of QT might fall into the narrative ``Challenges to traditional cybersecurity" as identified by \citet{Pohlmann2024}, or helping criminal organisations through the use of quantum communication devices that are virtually undetectable \citep{Vermaas2019}. Together, these ideas form the \textit{risk/benefit} frame.

\textit{Enigmatic frame}. Another area that received attention was instances where quantum 2.0 technology is framed as \textit{spooky/enigmatic}. In a discussion on making QT understandable to the general public, \citet{Vermaas2017} warned about the effects of  ``enigmatic" framing of QT: it can create an unnecessary barrier for stakeholders to discuss QT. \citet{Meinsma2023} found that in a quarter of the TEDx talks about quantum 2.0, this counterproductive framing was used. We aimed to provide a further analysis of this type of framing.

\textit{A focus on quantum computing}. Although QT can be subdivided into multiple technologies such as sensing, computing, and networks \citep{de_touzalin_quantum_2016}, TEDx speakers tended to centre their discussions of QT on\textit{ Quantum computing} \citep{Meinsma2023}. \citet{roberson2021talking} argued that a broader discussion of quantum computing is necessary for the realisation of the wider public good.

\textit{Everything is connected.} Finally, some TEDx speakers argued that quantum is a mystical concept, in which everything is interconnected through quantum fields. Such an approach can be seen as framing quantum science in a \textit{mystical/holistic} sense \citep{Meinsma2023}.

\subsubsection{Metaphors}
Metaphors are a common linguistic tool in making abstract concepts more accessible to a non-expert audience. Therefore, it is likely that this linguistic tool has become integrated into the communication and thinking of non-expert stakeholders. 

\citet{Lakoff1980} created a seminal theoretical foundation for the study of metaphors. When using a metaphor, one draws from a familiar context (source domain) in order to describe an abstract concept (target domain). In addition, they argue that metaphors can structure thinking and reasoning. 

Given that QST is built on unintuitive concepts, explanations often rely on metaphors \citet{Wackers2025}. For example, a content analysis of 816 Dutch newspaper articles by \citet{Wackers2025} showed that explanations of QT often use metaphors that draw from source domains such as ``Substances, materials and objects" (i.e. explaining superposition as a coin toss being heads and tails at the same time) or ``Entertainment, sports and games" (i.e. explaining wave-particle duality with the fact that music can be described using both melody and harmony). Source domains can be assigned through the use of the USAS semantic system, a framework set up to categorise all English words and some expressions into 26 different source domain groups, each denoted with a specific letter \citep{lancsUCRELSemantic}.

\subsubsection{Jargon}
Experts tend to use their own language of science, also known as jargon, which can be incomprehensible to non-experts  \citep{Krieger2016, Shulman2020}. However, the use of jargon is not specific to scientists, as other stakeholder groups such as business, government, and media have been found to use their own language and words \citep{Feigina2020, Richardson2008, Shuy1998-ow}. This language usage can make communication among different stakeholders more difficult. At the same time, jargon illustrates the way certain stakeholder groups think about and conceptualise quantum topics, with certain words holding specific importance within a field \citep{Hirst2003}. 

\subsection{Hype}\label{Hyping quantum technology?}
Stakeholders within QST can form expectations that are overtly positive (hype) or negative (alarmism) \citep{Dedehayir2016, Intemann2022-en}. For example, quantum scientists have been shown to engage in hype by exaggerating the expected benefits of their research to secure funding and interest \citep{SotoSanfiel2025}. Another example focused not on the potential benefit, but on the potential risks. \citet{mehnert_beyond_2025} suggested that the alarmist expectation that quantum computers could break cryptography could steer funding towards certain applications (for example, post-quantum cryptography). The exaggerated expectations created by hype (and alarmism) can not only promote funding, but also shape the direction of research and development in the field \citep{Roberson2023, disco_chapter_1998, borup_sociology_2006}.
Hype can be distinguished into two separate forms, named ``semantic" hype and ``discursive hype" \citep{Kari2023-cr_hypetypes}. The former describes an exaggerated or sensationalist tone in discourse, while the latter refers to a large wave of media coverage. These two types of hypes do not always occur simultaneously. Given that this work only gathers data from one-on-one interviews with stakeholders, only semantic hype can be studied.

Semantic hype has been shown to be prominent in communications and conceptualisations surrounding QST \citep{Meinsma2023, WangXu2026_typeshype, Meyer2023}. The other side of this phenomenon is alarmism, defined by \citet{Intemann2022-en} as statements that are ``inappropriately pessimistic, exaggerating risks and uncertainties" (p. 286). When a stakeholder describes the potential or future of QT using neither semantic hype nor alarmism, instead opting for a balanced picture, their rhetoric can be described as nuanced. This phenomenon is understudied in relevant literature regarding future visions of QT - only \citet{Meinsma2023} mentions that a marginal fraction of TEDx speakers in QST use a ``balanced view" to frame the benefits and risks of QT. 

To provide more structure to the study of hype, \citet{karami_quantum_2026} have created a concrete framework to assess statements about the future of quantum technology based on highly cited papers in QST. When discussing research/development on QT, scientists can describe certain ``use cases": particular applications or ways in which QT is predicted to be applied, such as breaking encryption or dealing with climate change. Furthermore, scientists can make a ``future projection": a conceptual goal or milestone in the development of QT, such as the realisation of a quantum computer or the development of a quantum algorithm. These use cases or future projections can be very vague (``faster computation"), highly specific (``simulating nitrogen fixation"), or anywhere in between. 

This degree of specificity can, but does not necessarily, reflect the level of semantic or discursive hype. Unspecific claims can easily fall into an exaggerated rhetoric, and might contribute to a high volume of media coverage, but this is not always the case. However, a more concrete assessment is that vague future predictions tend to be used by scientists in order to ``cast a wider net". Highly specific use cases can exclude potentially interested stakeholder groups, while a broader description of a potential technology can pique the interest of a large audience. When this desired engagement is reached, unspecific claims on the future of QT could foster discursive hype.

This study will apply the framework of \citet{karami_quantum_2026} to non-expert stakeholders to gain more insight into potential semantic hype in their communication style. In doing so, we keep in mind that the framework was designed based on scientific literature rather than spoken communication. However, we believe that using the framework within this different context will provide insight into how future depictions of QT in internal, non-expert stakeholder functions compare to public communication within scientific literature.

\subsection{Research questions}

This research project focuses on the use of frames, metaphors, jargon and hype by non-expert QT stakeholders, and therefore, we pose the following RQ:
\begin{center}
\textit{How can the communication by the non-expert QT stakeholders be understood through the lens of frames, language usage, and hype?}
\end{center}
\
This question is approached through the following sub-questions:

\begin{enumerate}
    \item \textit{Which frames, metaphors, and jargon do non-expert quantum stakeholders use when discussing and explaining quantum science and technology?}
    \item \textit{How do elements of hype (or alarmism) appear in discussions with non-expert quantum stakeholders, and how do framing and language use contribute to their prevalence? }
    \item \textit{In what way can framing and language use be understood through the desiderata of non-expert quantum stakeholders as described by \citet{Umbrello2024}?}
\end{enumerate}


\section{Methods}
\label{Methods}
To address the research questions, we performed 12 interviews with non-expert quantum stakeholders. Interviews allow for an in-depth inquiry into stakeholders' motives and desiderata, as well as a thorough analysis of their language use \citep{Chand2025}. We decided to conduct semi-structured interviews, as they combine both open-ended and closed questions, enabling a more structured analysis while still allowing broad input from stakeholders and allowing new frames to emerge \citep{AdeoyeOlatunde2021}. 

\subsection{Selection of participants}
For the criteria of ``non-expert" stakeholders, this study used the working definition presented in section \ref{subsection Non-expert stakeholders}. The relevant stakeholder groups consist of regulators and policymakers; investors and funders; industry partners and suppliers; ethics/societal experts; and media and journalists. These groups are selected for their unique role in the development of QT. That is, although they do not necessarily have formal training in quantum science, they have the ability to influence the development of QT. We left out two stakeholder groups: end-users and the general public. The end-users is left out because it is still unclear who belongs to this group. The general public is left out because it is too large and diverse to properly investigate through interviews.

We further sharpened our criteria of ``non-expert" to exclude stakeholders holding a master's degree related to QST.  This was done to ensure we sample stakeholders who do not have an extensive theoretical background with quantum concepts. Furthermore, learning quantum concepts outside academia leaves room for external influences to shape the intuition this study aims to explore, and stakeholders with a master's degree related to QST may already be well integrated into the world and language of academia.

The study opted to reach participants through non-random, purposive (snowball) sampling as this aligned with the project's capabilities and allowed us to use the network of contacts available to us \citep{Croucher2024}. To identify stakeholders, the authors' network was employed, as well as the Dutch national quantum ecosystem. Personal contacts, LinkedIn, and other available online information were also consulted, and when possible, stakeholders were asked to recommend others. These inquiries yielded a list of possible stakeholders and their contact information. The educational and occupational backgrounds of these stakeholders were reviewed to ensure they met the criteria. 

The selected stakeholders were contacted via email; in one case, a stakeholder was contacted in person at a quantum-related workshop. Stakeholders who did not respond to emails were called if relevant contact information was available. Twelve interviews were conducted, with at least two participants in each stakeholder group (see Table \ref{tab:stakeholders}).

\begin{table}[h]
    \centering
    \caption{Overview of the interviewed stakeholders. The stakeholder groups are based on the classifications of \citet{Umbrello2024}. }
\label{tab:stakeholders}
    \begin{tabular}{|c|c|c|}\hline
         \textbf{Stakeholder group}&  \textbf{Indicator}& \textbf{Area of work}\\
         Regulators and policy makers&  [Policy \#1]& National ministry\\
 Regulators and policy makers& [Policy \#2]&Governmental science institute\\
 Regulators and policy makers& [Policy \#3]&United Nations institute\\\hline
         Investors and funders&  [Funding \#1]& Public funding organisation\\
 Investors and funders& [Funding \#2]&Private funding organisation\\
 Investors and funders& [Funding \#3]&Public funding organisation\\\hline
         Industry partners and suppliers&  [Startup \#1]& Quantum research company\\
 Industry partners and suppliers& [Startup \#2]&Quantum research company\\\hline
 Ethics and societal experts& [Ethics \#1]&University\\
 Ethics and societal experts& [Ethics \#2]&University\\\hline
 Media and journalists& [Media \#1]&Financial newspaper\\ 
 Media and journalists& [Media \#2]&General newspaper\\ \hline
    \end{tabular}

\end{table}

\subsection{Interview setup}

The interviews took on average 55 minutes to complete, with interviews lasting between 45 and 66 minutes. The audio of the interviews was recorded with the participants' verbal consent. The interview protocol consisted of general questions that covered three main topics: stakeholder characterization, explaining quantum concepts, and visions on the future of quantum, which could be further elaborated on in the conversation (for the Dutch and English versions, see SI.2). These 3 parts specifically related to the research questions, as the first part related to the stakeholders desiderata and placement, the second part yields opportunities for specific QST topics to be discussed, and the last part relates to hype and stakeholder attitudes. Three pilot interviews were conducted to test and refine the interview protocol.

The interviews were preferably held in person at participants' workplaces to resemble the environment in which they were most likely to discuss quantum.  This was done to reduce the effect of being interviewed compared to a regular conversation \citep{Dawson2017}, and only one interview was conducted online. Furthermore, participants were asked to respond to questions and explain quantum concepts as if they were speaking with a colleague. The aim was for stakeholders to use the same language as in their daily office routine.

\subsection{Data analysis}
The interviews were conducted either in English or in Dutch. After each interview, the audio was transcribed using TurboScribe \citep{TurboScribe}. This tool was set to use the \textit{maximum accuracy} setting for two speakers (researcher and participant) and kept the original language. The resulting text was checked for transcription errors by consulting the audio recordings. The transcripts were analysed in their original language.

The transcripts were analysed via a directive qualitative content analysis \citep{Croucher2024}. Existing theory-based codes were used to code the transcripts, while also allowing new codes to emerge. The resulting codebook, including a description of all prevalent codes and how often they occurred, is provided in SI.3. Codes marked by an asterisk emerged during the coding process, as did all instances of jargon. The codes were grouped into the following overarching categories: Framing, Future visions, Jargon, Metaphors, QST discourse, QST ecosystem, and Stakeholder characterisation.

Codes for metaphors did not arise during coding but were all selected from the literature. Metaphors were categorised using the USAS semantic system, which has a specific library for both English and Dutch texts \citep{lancsUCRELSemantic}. As this system categorises the use of words within metaphors, multiple USAS categories can be applied to one metaphor.

Coding was performed with the ATLAS.ti software package \citep{AtlasTI}. The interviews and data analysis were divided between the two main researchers in this project. One researcher handled the interviews with participants from [Funding], [Media] and one from [Policy \#3], while the other researcher handled participants from [Ethics], [Startups] and two from [Policy \#1 and \#2]. During coding, both researchers discussed how to generally apply existing codes and possible emerging codes with each other.  

To assess intercoder reliability, the transcript of the [Policy \#3] participant was coded independently by both researchers. After coding was completed, the two transcripts were compared to analyse any differences between the codes. All differences were discussed, and consensus was reached on how each instance should be coded, as the differences were often pedantic or involved overlooked codes.

\section{Results}

This section covers all existing and emerging codes applied to the interviews, including their prevalence and examples of stakeholders using them. First, the context in which participants discuss quantum topics is briefly discussed. The second part contains codes relating to framing, metaphors and jargon. In the third part, all codes related to hype, such as alarmism, nuance, and scepticism, are described and connected to use cases and future projections. In the fourth part, we focus on how participants' backgrounds and desiderata relate to the previous three parts.

When participants are quoted, they are identified with their stakeholder group and numbered within square brackets, e.g. [Policy \#1] or [Ethics \#2] (see also Table \ref{tab:stakeholders}). Quotes in Dutch were manually translated into English by the first two authors. Although code frequencies do not serve as a direct result in this qualitative study, the differences in frequencies were taken as an indicator of the relative importance of the codes. The entire codebook can be found in SI.3.

\subsection{Context and daily reality of non-expert communication around QT}
The context in which participants speak about QST with each other helps us understand how language use permeates throughout these groups. This section discusses with whom and in what situations participants discuss QST and related topics, as directly asked during the interviews.

Overall, participants were found to discuss QST topics in both formal and informal settings. In formal settings, participants often had to discuss or explain quantum topics related to their work during meetings or presentations. In informal settings, such as during a break or an office party, or outside of work with friends and family, participants would bring up information about quantum as an interesting icebreaker. Furthermore, participants from the media and policy groups mentioned they could run stories on quantum by other colleagues, to make sure they were written/formatted in such a way that they are comprehensive to an audience that isn't informed on quantum, such as the general public or civil servants they have to communicate with. One participant mentioned how they would discuss quantum topics with colleagues, especially in an informal setting:

\begin{quote}
    [Media \#1]: ``I like to [...] perform a test run, it could be with colleagues at the coffee machine, and explain what my (quantum related) stories will be about. I like to see if it moves my unknowing colleagues, if they can understand it."
\end{quote}
The frequency and manner in which participants discuss QST vary widely across stakeholders. 

Some participants were the only person with knowledge of quantum concepts at their company. These people rarely speak with colleagues about quantum as it is often a less prominent topic at work, and when they do discuss it, it often consists of answering questions and explaining QST to others. participants at institutes that are more focused on quantum technology and its direct effects themselves, such as quantum startups, ethics experts, or funding organisations, often found themselves surrounded by people with more theoretical quantum knowledge. As a result, they often spoke about quantum by asking questions to experts present at their occupation, generally in informal settings, as these participants weren't always expected or required to have a high degree of specific quantum knowledge. The frequency at which people discussed quantum topics ranged from multiple times a day to once every few months, largely depending on how closely participants' occupations were directly intertwined with QT.

Participants can thus be found to discuss quantum topics, although the setting in which it is discussed and the people with whom they discuss it differ. Furthermore, although all participants are non-experts, their knowledge of quantum theory differs, especially relative to their colleagues, which affects how they discuss QST.


\subsection{Overview of framing and language use}

In the following section, framing and language use of non-expert QT stakeholders in the interviews is described in decreasing order of prevalence. The most dominant element in the interviews was the use of framing. Metaphors were slightly less pervasive, and jargon appeared only marginally.

\subsubsection{Framing}\label{results_framing}
We found several frames: Economic development, risk/benefit and social progress; Comparison to other technology; Ones and zeros, quantum computing; Difficult quantum phenomena in QST; Is quantum spooky, is it everywhere?; and several others. We describe these in the following.

\textit{Economic development, risk/benefit and social progress.} The dominant types of framing were of the categories \textit{economic competition/development} (54 instances), \textit{social progress} (46 instances), and \textit{risk/benefit} (56 instances) - all three occurring with approximately equal frequencies. These frames occurred primarily within the context of the future of QT, with some exceptions for the \textit{risk/benefit} frame. 

For instance,  the \textit{economic competition/development} frame tended to involve statements around a ``quantum race" or large investments and collaborations, examples being:
\begin{quote}
    [Policy \#1]: ``Especially with the quantum computer, there is a competition about which type of quantum computer will become dominant. Well, throwing large sums of money towards it always helps. People will start creating their own types of quantum computers for specific applications. That could be startups, but most likely it will be done by multinationals. [..] Organisations will have to collaborate and pay the price for what [QT] can deliver."
\end{quote}
\begin{quote}
    [Media \#1]: ``Quantum computing could become an international line of business where Dutch corporations would have achieved a good advantage. That is something for us to pay attention to. Investments going into this field will reach billions."
\end{quote}

The approach to the risks and benefits of QT can be divided into two categories, namely through the context of (cyber)technology and through a lens of the wider public good. The risks and benefits regarding (cyber)technology mostly revolved around cybersecurity and secure communications, and were coded through the \textit{risk/benefit} frame. These technical risks/benefits were mentioned when asked to explain concepts from QST, as well as in the future visions section of the interview. As an example of the former, when explaining the quantum computer, one participant stated that it can ``simulate natural processes a lot better than current computers can" [Ethics \#1]. When asked to explain quantum internet, [Startup \#2] envisioned internet searches to become ``instantaneous" because of added speed from ``quantum processes".

When asked what the future of QT will be, participants often pointed to the importance of adequate preparations. They typically argue that since quantum computers and communications might (or will) introduce a paradigm shift in cybersecurity, current systems need to be updated. A typical example when asked about the potential impact of quantum computing:
\begin{quote}
    [Funding \#1]: ``Privacy is another important domain. What does that mean? Generally with our internet activity, what is being tracked? And who uses that? Well, with quantum computers the risks will get a lot more serious. I think this could impact the banking sector as well. All this encryption and decryption will need to get much more secure."
\end{quote}

The other side of risks and benefits is the context of the wider public good, and the \textit{social progress} frame was applied there. Benefit-wise, this frame often came up when asked about the future impact of QT. When prompted in this way, participants mentioned healthcare benefits, scientific progress, or the possibility of achieving more sustainable energy technology. Some participants expressed a more general, hopeful sense that QT can bring progress to society as a whole. One participant [Startup \#1] stated this sentiment when asked about their motivation to work in QT. 

Risk-wise, most participants focused on ethical and policy-related concerns. These issues were typically discussed when asked about the future of QT. These concerns typically revolved around increased inequality, where only a few powerful organisations truly profit:

\begin{quote}
[Ethics \#1]: ``Maybe quantum will be a revolution for Jeff Bezos' pockets, you know, or whoever is leading these companies. They'll probably get really rich. But I think for the average person, probably not."
\end{quote}

\textit{Comparison to other technology.} Two new codes that emerged during analysis were \textit{comparisons with traditional/classical technology} (36 instances) and \textit{comparisons with Artificial Intelligence} (AI) (34 instances). When asked to explain quantum technologies, participants often built their narrative around how current technologies work - here, current technologies were often described as ``classical" or ``traditional". The level of detail regarding the relation to traditional technology differed between participants. Additionally, we saw how, in some instances, participants framed QT as generally ``better" than traditional technology (13 instances). These observations relate to the different levels of specificity in use cases, which will be discussed further in Section \ref{hype}. 

Moving on, when asked to describe how the future of QT will unfold, most participants had a tendency to contextualise the impact and timeline of QT through a historical lens - the industrial revolution was mentioned, for instance. On multiple occasions, participants referred to the development and implementation of AI. Note that historical comparisons and the mention of AI were not specifically probed for by the interviewer. One participant [Funding \#1] predicted the general public would develop QT literacy comparable to how the public developed AI literacy. 

Participants often compared either the revolutionary possibilities or limitations of QT to the development of AI, some examples being:
\begin{quote}
    [Ethics \#2]: ``I think AI is truly a general purpose technology that will bring about drastic changes in how we do things. Throughout all of society, from the breadth of our private lives to our working lives. And I just don't think that will be the same for QT,  although QT could catalyse important scientific developments."
\end{quote}
It is also worth noting that the participants used AI to illustrate the uncertainties in dealing with emerging technology. As one participant commented that
\begin{quote}
    [Policy \#1]: ``You always see that new technology finds its place, it might be in ways you expect, but also in ways you don't expect. [\textit{Describes large scale survey on AI usage.}] I don't think anyone would have expected that AI would become a therapist to some, or to others a best friend or for others a relationship. Especially not at this scale."
\end{quote}

\textit{Ones and zeros, quantum computing.} The concepts of superposition, entanglement, and quantum computing were frequently explained through \textit{ones and zeros} - referring to the binary system computers use to perform any task or calculation (22 instances). For example, most participants explained superposition as ``being one and zero at the same time" or ``being between one and zero". Secondly, most participants tended to frame concepts from QST in terms of \textit{quantum computing} (22 instances). For instance, when prompted to explain superposition, a participant might immediately mention that a quantum computer performs calculations by using qubits, which can be one and zero at the same time (showing overlap between the two frames). This focus on quantum computing also appeared when participants were asked about the future of QT: participants often described the future of quantum computing in particular detail compared to other aspects of QT. 

However, participants did mention other quantum technologies, sometimes in significant detail. Some participants named multiple use cases of quantum sensing when explaining the technology. Without further prompting, two of these participants remarked that quantum sensing is closest to realisation compared to other applications of quantum science.

Another technology that received reasonable attention was quantum communication. When we asked about quantum internet, three participants gave a detailed explanation of the security in quantum internet while most of the rest seemed to have a basic understanding of the security principles. Additionally, three participants were familiar with the term Quantum Key Distribution (QKD). All in all, the partcipants seems to be familiar with there three types of quantum technologies. Although quantum computing is the one that they most familiar with.

\textit{Difficult quantum phenomena in QST.} Some participants framed QST as being \textit{complex/too difficult} (20 instances). When asked about the general future of QST, one participant [Funding \#1] said that QST is very ``specialised" and would be difficult to understand for the general public. Based on this statement, the participant argued that knowledge about quantum would stay within development circles. One person [Media \#1] called QST ``hypertechnical" and not typically interesting to the general public, after inquiry about their daily communications around QST. In some cases, participants framed the complexities of QST as \textit{outside of their scope/irrelevant to their job}. 

In a similar vein, certain participants explained or referred to QST in terms of ``quantum phenomena" or "qualities/traits of quantum mechanics" - typically when explaining concepts from QST. In some cases, these participants would then clarify how these quantum phenomena worked, and the mention of ``quantum phenomena" thus signals further explanation would follow. 

However, in other cases, participants mentioned ``quantum phenomena" without providing further explanation. In some cases, the term ``quantum phenomena" was not said explicitly, but a similar allusion was made, which we coded under this frame. Two examples are shown below, one where the term ``quantum phenomena" is explicit and one where there is only an implicit mention. 
\begin{quote}
    [Ethics \#2]: ``The quantum internet is a form of communication where you use quantum phenomena. I think it always constitutes a connection between two quantum computers. You would let quantum computers exchange information not in the classical way, but in the quantum way."
\end{quote}
\begin{quote}
    [Policy \#1]: ``Quantum sensors are sensors based on quantum mechanics. They can distinguish certain particles more specifically, thanks to the qualities of quantum mechanics."
\end{quote}

\textit{Is quantum spooky, is it everywhere?} Among non-expert stakeholders, the frames spooky/enigmatic (12 instances) and mystical/holistic (4 instances) were rare, often occurring only as remarks in addition to broader discussion. For example, when asked to explain quantum entanglement, [Policy \#2] started by characterizing it as ``one of the most enigmatic phenomena in the quantum world". They did not touch upon this statement any further in the rest of their answer. The \textit{mystical/holistic} frame only appeared in one interview: when explaining quantum concepts, a participant [Startup \#2] described how in the universe everything is interconnected through quantum entanglement. 

The \textit{spooky/enigmatic} frame appeared in four interviews, usually in the form of a brief mention such as labelling something ``enigmatic" or as ``spooky nonsense".  One participant [Media \#2] discussed the fact that quantum can be an enigmatic concept to people who are unfamiliar with QST. Below, an example of a more in-depth discussion is shown:
\begin{quote}
    [Media \#2]: ``I had a long discussion with my editor in chief at the time. He thought that the Delft experiment of entanglement just shouldn't be right. That it couldn't be possible. But I thought that that was exactly what made it so valuable. It is possible, it has been measured. This does not fit with any logic. But sure, that's possible. You can put a disclaimer with it. I actually always love that: '\textit{I am so sorry, this will probably hurt, but this is what was actually measured.}' But I don't think you should dumb it down." 
\end{quote}

\textit{Other, notable types of framing.} The interviews yielded four new, but minor types of framing. Two revolve around the everyday use of QT, namely whether QT will become \textit{available for personal use} (9 instances), and the fact that QT is already \textit{present in everyday life} - LEDs, MRI (12 instances). Moving on, two participants explained QST through the figures \textit{Alice and Bob}: these names stem from traditional explanations of physics phenomena. Finally, some participants made sense of the development of QT through a\textit{ historical perspective on other technologies} (11 instances). Examples are mentioning the history of previous revolutions, or how QT will render traditional cybersecurity outdated, just like in the history of military innovations. Because this type of framing tends to co-occur with future predictions, more examples of historical perspectives will be discussed in Section \ref{nuance}.

\subsubsection{Metaphors}
Metaphor usage was common, occurring at least once in every interview. They were mainly applied when participants explained QST concepts, but participants were also asked if they knew and could give an example of specific quantum metaphors. To categorise the metaphors, the USAS semantic system was applied \citep{lancsUCRELSemantic}. Although our analysis was qualitative, it became clear that one semantic category was significantly more prevalent than others: Movement, location, travel and transport (M) (31 instances). For example, participants might describe superposition as a particle being ``between" states, or with a coin that is ``spinning" in the air. Another participant described that calculations by a quantum computer are similar to ``walking" all paths of a labyrinth at the same time. Other dominant categories were Social actions, states and processes (S) (15 instances) and Substances, materials, objects and equipment (O) (13 instances). Almost all other semantic categories occurred at least once. It should be noted that most instances of metaphor use involved multiple semantic categories at the same time.

Generally, metaphor usage was very diverse. The well-known metaphors of ``Schrödinger's cat" and ``a coin spinning in the air" were mentioned more than once, though the first was subject to some debate: one participant thought it was a great metaphor, but another believed it did not make any sense. The general use of metaphors was also discussed in both a positive and negative light. For instance, one participant [Policy \#3] aims to use metaphors wherever possible and aims to always keep optimising in this regard. On the other hand, one participant [Media \#2] generally avoids metaphors and argues that journalists should not get ``too poetic". 

Overall, most participants mentioned at least one unique metaphor. In table \ref{tab:metaphor_table},  metaphors we deemed uncommon based on literature or prevalence in interviews are shown by source domain and target domain, together with the stakeholder group of the participant and the relevant semantic tags. One metaphor deserves a separate mention: A participant [Policy \#3] described how they once explained quantum entanglement to a group of young children by playing a game. In this game, the children were paired up and made to face each other. The game was to close your eyes and raise either the left or right hand. When one child raised their left hand and the other raised their right, the kids were entangled. These hands-on explanations of QST to children is a routine part of this participant's job.

Lastly, we discuss comparisons to other technologies in the sections on framing, but within this comparison also lies a metaphor. An important occurrence is when participants explained the quantum computer in terms of how our current computers work. This is a metaphor in itself, since quantum computing is fundamentally different from ``classical" computing \citep{quantum_computing_vs_classical}. 

\begin{table}
    \centering
    \setlength{\arrayrulewidth}{0.5pt}
\caption{Novel metaphors when explaining/discussing QST. From left to right, the columns show the stakeholder group of the participant mentioning the metaphor, the target concept that the participant wanted to explain, a description of the metaphor, and the source domain according to USAS semantic categories \cite{lancsUCRELSemantic}.}
\label{tab:metaphor_table}
    \begin{tabular}{| c  |c  |>{\raggedright\arraybackslash}p{0.25\linewidth}|>{\raggedright\arraybackslash}p{0.25\linewidth}|}
        \hline
        \textbf{Group} & \textbf{Target concept} & \textbf{Metaphor} &\textbf{Source Domain}\\\hline
        Startup & Superposition & A proposal draft before the final signature &general and abstract terms; money and commerce in industry; languange and communication; time\\
        \hline
        Startup & Entanglement & Linked Excel cells &numbers and measurements; science and technology\\
        \hline
        Startup & Superposition & ``Sinterklaas" being everywhere at the same time&movement, location, travel and transport; names and grammar\\
        \hline
        Policy & Entanglement & The ``L" and ``R" markings on socks&the body and the individual; languange and communication\\
        \hline
        Policy & Superposition & Kid playing with a light switch, by flickering it on and off at high speeds &entertaiment, sports, and games; movement, location, travel and transport; world and environtment\\
        \hline
        Funding & Quantum particle & Waves after throwing a rock into a lake &movement, location, travel and transport; substances; materials objects and equipment; world and environtment\\
        \hline
        Media & Quantum computing & Letting an entire soccer team, including sidelined players, try to score a goal at the same time &entertaiment, sports, and games\\
        \hline
    \end{tabular}

\end{table}

\subsubsection{Jargon}\label{results_jargon}
Jargon played a minor role and was used only sparingly by most of the participants. The participants rarely integrated jargon into their explanations or discussions, and most terms were used only once per stakeholder. Jargon was used evenly across stakeholder groups. The most frequent occurrence of jargon was in the form of ``Quantum" $+$ ``some terms" (35 instances). In most cases, such terms were only mentioned after being asked whether there are important quantum concepts that were not discussed before. Examples are: ``Quantum advantage", ``Quantum divide", ``Quantum awareness",  ``Quantum Inspired", ``Quantum space". Most of such terms were mentioned once, but the notion of a ``Quantum space" was mentioned more frequently by one participant from the regulator and policymaker group (10 instances), and the term ``Quantum inspired" was used by 3 participants (4 instances). Other instances of jargon include Moore's Law (in a prompted discussion about the innovative value of QT), QKD, and Post Quantum Cryptography (PQC) (both in a prompted explanation or discussion of the quantum internet and its safety concerns).

Another theme we identified was that some participants showed surprisingly niche, technical knowledge (23 instances). The knowledge is deemed `surprising' because, to the authors of this paper, this level of abstraction did not seem to be strictly necessary for their job. All three participants from the policy category seemed to show sophisticated technical knowledge, as did three others from funding, media, and startups. It appeared in varying intensities throughout more than half of the interviews, and sometimes co-occurred with jargon. Examples are very diverse, but some common instances involved linear versus exponential scaling of computation, detailed knowledge of the hardware of a quantum computer or a quantum chip, and intricacies of quantum communication protocols. For example, in an attempt to explain how a quantum computer might work, he said:

\begin{quote}
[Policy \#1]: ``We can think of experiments we did, with quantum algorithms, that were quantum-inspired, hybrid. That means that you use a HPC, a high-performance computer. You can connect it to a quantum computer and already do some calculations with it, as preparatory work for the quantum computer." 
\end{quote}

\subsection{The language of future visions}

\subsubsection{Hype and alarmism} \label{hype}
As hype and alarmism seem to be two opposites that both relate to exaggerations of attitudes towards new technologies, they are discussed here together \citep{Intemann2022-en}. Hype is present when participants exaggerate the beneficial sides of science in an inappropriate or incorrect manner; alarmism is present when risks and uncertainties are exaggerated, showcasing inappropriate pessimism towards science \citep{Intemann2022-en}. The prevalence of hype and alarmism during the interviews was lower than expected based on the literature. Hype was present in 7 interviews (12 instances) and alarmism in 5 (8 instances), and often only a single instance per interview. The usage of hype and alarmism did not seem to be linked, and they did not occur at the same time.
When hype occurred on its own (as opposed to being part of a future vision), it was mostly an expression of the impact QT would have on society. However, even here the claims are not extensive. One illustrative example of hype is:

\begin{quote}
[Media \#2]: ``I would guess that quantum computing is on a scale of the internet. Minimum. It could even become much larger."
\end{quote}

It is important to note that such exclamations were often part of a more nuanced discussion of QST, as discussed in the following section. Alarmism was most prominent in discussions surrounding security, shown by overlap between alarmism and mentions of encryption and the \textit{risk/benefit} frame. Three participants mentioned being scared about the future of quantum. Although alarmism was generally only briefly mentioned (if at all), one participant [Startup \#2] mentioned it for a longer segment of the interview. When discussing the future of QT, a participant said:

\begin{quote}
[Startup \#2]: ``I have the fear, as I previously stated, that when the computational capacity is connected everything will be done by AI, and everything will be decided for you. From how you will handle affairs, to what you can spend your money on, and the sorts. A digital prison."
\end{quote}

\subsubsection{Nuanced views}
\label{nuance}

Nuance, in this context, entails any balanced account of risks and benefits regarding QT. Nuanced views were common throughout all interviews when discussing the future and potential of QST. These discussions were based on balance instead of hype or alarmism. Although our content analysis was qualitative in nature, we marked nuance about four times more than hype and alarmism combined (87 instances compared to 20). Participants were reserved in making definitive claims about the future of QST: when prompted to sketch their vision of the future, participants often expressed that it is ``impossible to predict", or even ``a pointless question".  In the instances where participants did explain their visions of the future, it was generally through a balanced reflection of current affairs of the quantum ecosystem. 

In six interviews, there were instances where nuance was approached purely through \textit{economic competition/development} framing. In such cases, the participants focused on the technical challenges in QT development. This often came with moderate to high degrees of technical knowledge (Section \ref{results_jargon}). Another form of technical nuance occurred when participants expressed that they don't expect QT to become a general purpose technology. Often comparing to AI (also mentioned in Section \ref{results_framing}), participants viewed QT as an ``enhancement" as opposed to a society-wide revolution. Here, one participant used a metaphor for this concept:
\begin{quote}
    [Policy \#1]: ``I am not sure if QT is similar to [digital revolution]. I don't think it's like the difference between a broom and a vacuum cleaner. It's more like having a new attachment to the vacuum cleaner, so that you can reach the nooks and crannies that you couldn't before."
\end{quote}

Six participants approached nuance through the \textit{social progress} frame, by stressing that the merits of QT depend on policy and the intent of who employs the technology. One participant used a metaphor to illustrate the importance of regulations around QT by comparing QT to a knife, which can be used for food preparation and to harm others. This framing through \textit{social progress} was sometimes accompanied by either \textit{risk/benefit} framing or \textit{economic competition/development} framing. For example, one participant [Policy \#1] explained three increasingly alarmist scenarios for the Netherlands because of decreasing government investment in quantum technologies. Another participant [Ethics \#1] mentioned how government spending will shape the development of quantum:

\begin{quote}
    [Ethics \#1]: ``Whatever happens with quantum technologies is being shaped today. It's being built today by the laws, by the regulations that we're building around it. It's being built by the quantum national strategies that countries are writing because all of that is determining where the money is going, what research avenues are being prioritized, and what future applications we will have."
\end{quote}

Other instances of framing when creating a nuanced view of QT occurred through a historical perspective, sometimes combined with a \textit{comparison to AI}. An example of the latter is mentioned in Section \ref{results_framing}, where a participant [Policy \#1] discussed the unexpected ways in which AI has found its way into society. Generally, historical framing was typically prompted when asked about a possible ``quantum revolution". Two more examples are:
\begin{quote}
    [Funding \#1]: ``The word revolution is associated with something that changes instantly. But [historically], revolutions take  a long time. [..] I do think that a revolution means that something irreversibly changes in society. I also see quantum in that way, that it could really bring about a revolution."
\end{quote}
\begin{quote}
    [Policy \#1]: ``We have always had revolutions - industrial, digital and all those things. Every time, it has found its place in society and we move on. Right now, our generation is a bit in between revolutions. We are already quite used to the digital revolution, and already start using less social media than our parents. [..] I am not sure whether quantum can create an entirely different world."
\end{quote}

\subsubsection{Scepticism}
\label{results_scepticism}
Another future vision shared by multiple participants was one of scepticism (17 instances). Here, this type of future vision is defined as doubts about any claims made regarding QT. Scepticism was present in six interviews, and occurred only briefly. Most scepticism centred around the future of quantum. Typically, participants framed their scepticism in the context of \textit{economic competition/development}: they were sceptical of grand claims made by industry actors, and expressed doubts about the actual use cases of QT. Some comparisons were made to empty promises in the AI industry. Examples of scepticism through the frame of \textit{economic competition/development}:
\begin{quote}
    [Startup \#2]: ``When there is this much money involved, they will want things too much. We've already had some papers from our group that had to be retracted, you must have also heard the story. You want to be able to show investors some value, and sometimes you want to show things too early before they are true. So then you get a sort of lying in science, but only to please the investors."
\end{quote}
\begin{quote}
    [Ethics \#2]: ``As far as I know, people [in QT] are making huge leaps in their claims between where we are now with quantum computing, and what we could ultimately do. There are a lot of steps between those things, a lot of assumptions."
\end{quote}

In some instances, participants added nuance to their own scepticism by stressing that they are not an expert on QT. One participant explained that they don't have a tangible reference to imagine the applications of QT (this also involved a historical perspective), and another described their experience as a non-expert in a highly technical QT startup:
\begin{quote}
    [Policy \#3]: ``I struggle with it because it's very hard for me to imagine what that's going to look like. Because cartoons growing up, they kind of had this element of the future, so self- driving cars in movies were things, or robots. There were all these tangible things that people imagined would be the future of flying cars and such. And we do have self-driving cars, right? Maybe not flying cars yet, but there are people doing this. So there was a way to kind of imagine what that kind of future would look like. With quantum, there's no relatable reference that somebody actually put it in a way, that could translate that way."
\end{quote}
\begin{quote}
    [Startup \#2]: ``So my knowledge is a little sparse, but to me all of this here seems like a flea circus. There is a tamer in front of a crowd, with a hoop. He says 'Little Mary is jumping through the hoop!', and nobody really saw it, but everybody starts applauding and throwing money. And that is my cynical view on all of this. I can't really control what happens [at the nanoscale of quantum]."
\end{quote}

Finally, two participants were sceptical of jargon terms such as ``quantum revolution" and ``quantum internet". One participant  [Media \#2] expressed that the notion of a quantum revolution is ``just a marketing term", and another one [Policy \#3] explained how people talk about a quantum internet while that ``does not even exist". In the latter, the participant explained in thorough detail what ``internet" actually means, and how quantum communication is not there yet.

\subsubsection{The presence of future projections and use cases.}
The framework of future projections and use cases, as described in section \ref{Hyping quantum technology?}, allows for a more advanced analysis of possible occurrences of hype and to further the understanding of future visions. Use cases were present in all 12 interviews and were coded in 88 instances. Notably, non-specific use cases (49 instances) were more prevalent than both rather specific (16 instances) and specific (23 instances) use cases. Importantly, although every participant gave at least one non-specific use case, only 7 participants gave rather specific use cases, and 8 gave specific use cases. Use cases were mostly mentioned when explaining quantum concepts or when discussing the impact and future effects of QST. They came up naturally during the interviews, and participants were not asked directly when or if they would normally mention them on a day-to-day basis. Participants did not mention use cases in the first section of the interview on participant identification.

Use cases focused mostly on 3 prevalent quantum technologies: quantum computing, quantum internet and quantum sensing. These were also the three technologies which the participants were prompted to explain during the interview. The most commonly mentioned use cases centred on quantum computing, which was most often mentioned when the stakeholder was tasked with explaining the relevant quantum technologies. In decreasing order of prevalence, discussed use cases were: simulating molecules or materials, replacing data centres, making weather predictions, improving logistics, faster health analyses, helping businesses, and trading/finance. Certain participants mentioned that quantum computers were only usable for very specific tasks and not broadly applicable. This contrasted with other stakeholders who were unsure about possible use cases, but mentioned that they might be usable for everything, even personal use. Although less common, use cases for quantum sensing were also mentioned in the form of improving GPS/navigation, allowing underground pipelines to be registered, or simply performing [Policy \#2] ``more precise measurements". Use cases for the quantum internet were also less common, and ranged from being more secure and applicable for banking to simply [Startup \#1] ``a faster internet". Regarding QT in general, three participants mentioned that use cases were still unknown and would become clear in the future.

When non-specific use cases were mentioned, participants often also mentioned that they were unsure about the related technology and its applications. Furthermore, when participants mentioned non-specific use cases, they looked to current technology and treated a new quantum technology as a direct upgrade (for example, a quantum computer being better at computer tasks, a quantum sensor being better at sensor tasks). Illustrative are the following participant examples, the first of which was mentioned when the participant was asked which problems could be solved in the future by quantum and the second when asked to explain the quantum internet:
\begin{quote}
[Funding \#1] ``For everything which you would currently use the calculating power of a computer,  (...), you would be able to use a quantum computer."
\end{quote}
\begin{quote}
[Startup \#1] ``In daily life we use internet with the classical computers to get information from one place to the other. The same is true with the quantum Internet. The information will travel incredibly fast, so you don't have to wait seconds, it will just be there."
\end{quote}

When participants mentioned specific use cases, they did not claim certainty about this technology. The specific use cases were often present as an addition to the non-specific use cases, elaborating on a previous answer. For instance, when discussing problems QT could solve, one participant mentioned:

\begin{quote}
[Startup \#1] `` I'm also thinking in medicine it could invent a new medicine that would help to save a lot of lives. It will be able to process information, like I said earlier, at a much higher speed. For example, you go give a blood test, you don't have to wait days to find out the results. You'll be able to find right then anything."
\end{quote}
Although this quote starts with a non-specific use case (``new medicine that could save a lot of lives"), it then goes on to a specific use case for QT (an instant blood test).

Future projections, such as goals and milestones (both 17 instances), were much less prevalent in the interviews. Most instances focused on completing a functional quantum computer, including milestones to reach this goal, such as developing software for quantum algorithms. These projections also mentioned that it would take a while before a quantum computer could be realized. The participants would often mention such projections when directly asked about the future (of quantum technology), but also when tasked to explain what a quantum computer is to a colleague. 

Another milestone which came up focused on the wider social requirements for the development of QST, with some predicting an influx of funding, business, research, and standards, as a pre-requisite for realizing QT. Participants often brought up these issues during a general explanation of the effects and workings of QST.

\subsubsection{Framing and language use for use cases and future projections.}
The prevalence of certain use cases coincided with framing and language use. The most common frame often present with these use cases was \textit{comparison to other technologies}. Here, stakeholders often looked at previous uses of existing technologies to predict the future of quantum technologies. One instance was for quantum sensors, where it was explained that, similar to how traditional sensors measure different information, quantum sensors can do so more accurately as they are more advanced. 

The \textit{risk/benefit} frame also overlapped noticeably with use cases. Here, QT was often described via beneficial use cases, such as improving weather predictions or use in the logistical sector. Risks were present in the form of malicious use cases. When asked about how the future could change through quantum (without asking if it would be positive or negative), one participant cautioned how:
\begin{quote}
[Media \#1] ``Quantum can have a gigantic impact if the quantum computer can break all data-encryption. Without anyone realising that it occurs. And without having created proper defences. In such a case, many things that keep society together will be broken down."
\end{quote}

In a similar vein, the frame of \textit{social progress} also saw significant overlap with mostly positive use cases. Furthermore, as will be discussed in more detail later,  when use cases are mentioned, they are often shown in a nuanced way, with both possible benefits and risks. As one stakeholder answered when asked what problems could be solved with quantum:
\begin{quote}
[Media \#1] ``Perhaps that trains will never be late, and that the NS (public railway) timetable will always be correct. Perhaps that a global problem will be solved, so that no one will have to be hungry again. But I do not see anything of this sort happening. A lot has been predicted about what the (quantum) computer can do for humanity.  This could be possible. It mostly seems like a very large promise."
\end{quote}

No general trend can be found between the use of specific or non-specific use cases and the previously mentioned frames, although a trend does seem to exist for individual participants. One frame that had significant overlap with non-specific use cases was \textit{because of quantum}. Here, people often claim that using ``quantum" allows for non-specific applications of new QT. This frame coincides with explanations of QST, as shown here where a participant was asked to explain what quantum sensors are:
\begin{quote}
[Policy \#1] ``Quantum sensors are sensors based on quantum mechanics. They can distinguish certain particles more specifically, thanks to the qualities of quantum mechanics.. (...). This allows you to perform very specific measurements you can't do with regular sensors." 
\end{quote}

The only clear links to language use could be found with specific use cases. These were found to overlap more with a high degree of technical knowledge and the use of jargon. It seems to be that in-depth knowledge of QT allowed for discussion of very specific use cases. For example, one participant [Policy \#3] used jargon such as PQC and QKD to show the possible applications of quantum communications and the quantum internet. The use of metaphors for use cases was rare, occurring only once.

\subsection{Framing and language use understood through stakeholders' desiderata}

In this section, we analysed the stakeholders' language through the lens of their desiderata. Our aim was to investigate whether the framing and language use above can be explained using stakeholders' desiderata.

\subsubsection{Stakeholder background and knowledge}
The participants were asked about their (educational) background. Although none of the participants was a quantum expert, they are relatively highly educated; 8 possessed a master's degree, and only 1 had not obtained a bachelor's degree. Although these degrees did not relate directly to QST, they did relate to fields adjacent to QST, for instance, engineering or computer science.

Despite their educational level, participants mentioned that they did have knowledge of QST, often obtained through autodidactic endeavours (reading related literature, watching videos about quantum topics) or by simply conversing with colleagues. However, this largely differed per participant. Where one stakeholder [Policy \#3] mentioned they would ask as many questions to expert colleagues as possible, another [Media \#2] explained that quantum was seldom discussed or explained in their field. This can be related to the fact that many participants are only occupied with quantum part-time, and thus don't have enough reason to discuss it with colleagues. As two participants put it:
\begin{quote}
    [Media \#1]: ``I do much more than just quantum, and there isn't that much going on with it. We aren't a newspaper which specifically writes for the quantum world."
\end{quote}
\begin{quote}
    [Funding \#1]: ``I don't work on this (quantum) topic on a daily basis. I think its more like once per two weeks"
\end{quote}
\subsubsection{Link between stakeholders and their desiderata}
The participants were asked to describe their work and motivation, which are linked here to their stakeholder group identification as outlined by \citet{Umbrello2024} (see Table \ref{tab:stakeholder desiderata table}). Based on these descriptions, we draw a connection to desiderata we would expect stakeholders to have. In the following, we show how stakeholder groups relate to the desiderata we base on the work of \cite{Umbrello2024}.

The participants in the regulators and policy segment interviewed in this research mainly focused on developing regulations and policies for (the integration of) future QT. As expected from their desiderata, these participants looked at the societal implications of QT, as mentioned by one participant who explained their job: 
\begin{quote}
    [Policy \#3]: ``So how to make sure that whatever is deployed, whatever advancements are made in the quantum space do not make things worse when it comes to access."
\end{quote}

The motivations of participants in the investors and funding segment also overlapped with their desiderata as described by \cite{Umbrello2024}, acknowledging their importance in directing and affecting the future of quantum developments. However, these participants generally did not mention funding itself, nor did they focus on commercialisation.  

The group of industry partners and suppliers described their tasks and motivation as ensuring that the startups were operational, both from a maintenance and communication point of view. Similar to the previous participants, they also mentioned the effect quantum developments could have on humanity.

Interestingly, one participant from the industry partners and suppliers group did not align with their expected desiderata, as they did not care about the future of quantum developments given their background. When asked about the impact of quantum on their line of work, the participant focused on the importance of the technology for their own livelihood, stating:
\begin{quote}
    [Startup \#2]: ``If it were my choice, the quantum computer would never be realised. Because when the related research is finished, I am also finished. Then I can go and look for a different job."
\end{quote}

Participants in the ethics group focused largely on the ethical and social aspects of QT, mentioning a variety of possible ethical issues regarding QST. Furthermore, they clearly stated their interest in addressing current quantum developments and ensuring that they could be adequately accounted for. As one participant put it when asked about their goals and motivations:
\begin{quote}
[Ethics \#1] ``The goal of what I do and my work is to try to encourage more responsible, more ethical development of these technologies so that in the future, yeah, we can all enjoy them more and don't have to deal with their negative impacts that much."
 \end{quote}

For the stakeholder group of media and journalists, informing the public on developments within quantum was found to be important. They mentioned the importance of being critical of the information they obtained.  One stakeholder, who worked for a finance-focused newspaper, mentioned how they only discuss quantum topics when they are related to finances, as that would be most relevant to their audience. This focus on relevance to a wide public/their own audience became clear when a participant discussed how the topic of "quantum" was not a daily occurrence for them:
\begin{quote}
    [Media \#1]: ``When we write about quantum, it has to be interesting for the readers. (...) for such a hyper technical subject as quantum, it has to be interesting for as many people as possible, because we don't write for the limited number of real quantum scientists."
\end{quote}

Although the participants do not always directly state their expected interests and motivations, the desiderata of the participants roughly align with the framework set out by \cite{Umbrello2024}.
\begin{table}
    \centering
    \caption{Stakeholders and their expected desiderata. The desiderata were based on stakeholder descriptions from section 3.3 and the reasoning column of Table 2  \citep{Umbrello2024}, as these most corresponded to the concept of desiderata. }
    \begin{tabular}{|c|>{\centering\arraybackslash}p{0.6\linewidth}|}\hline
        \textbf{ Stakeholder groups}& \textbf{Desiderata and ethical considerations}\\\hline
         Regulators and policy makers& ``While they aim to foster innovation, they must also protect citizens from potential harm. Ethically, their position is often a balancing act between facilitating technological growth and ensuring societal safety and equity (p.11)" 
and 
``[...] They have the responsibility [and interest] to ensure that QT is developed and used in a way that aligns with society’s values and
interests, protecting public welfare and promoting RI (Table 2)"\\\hline
         Investors and funders&  ``[...] they have an interest in ensuring that the technology is developed responsibly and provides a return on investment that aligns with societal values (Table 2)"\\\hline
         Industry partners and suppliers& ``These stakeholders primarily aim for technological advancement and commercial viability. However, their interests might sometimes prioritize profit over broader societal benefits, leading to ethical concerns like data privacy, monopolization, or the rapid deployment of under-tested technologies. While industry players are essential for technological advancement, their primary
focus on commercial viability may at times conflict with the broader societal values, such as privacy and equity. This tension highlights a potential area of conflict, particularly when profit motives might overshadow ethical considerations." and ``Their involvement in the QT ecosystem can influence the design, availability and adoption of the technology, and their values and interests should be considered in the development process"\\\hline
         Ethics and societal experts& ``They help identify, analyze and address potential ethical,
social and legal implications of QT, ensuring that the
technology is developed and used in a manner consistent with societal values (Table 2)"\\\hline
         Media and journalists& ``They play a crucial role in
shaping public opinion,
raising concerns and
generating constructive
discussions that can guide
the development and use of
QT in a way that aligns with
societal values. By accurately
and responsibly reporting on
QT, they help ensure that
the wider public stays
informed and engaged with
the technology’s progress and
ethical considerations"\\ \hline
    \end{tabular}
    
    \label{tab:stakeholder desiderata table}
\end{table}

\subsubsection{\textit{Relation between framing, language use and stakeholder desiderata.}}
After we established that our participants seem to align with their desiderata, we investigated whether certain frames are related to certain stakeholder group and their desiderata. For example, investors and funders might be expected to use the \textit{economic competition/development} frame since their desideratum is about return of investment.

We found that the use of the three most dominant frames, \textit{economic competition/development}, \textit{risk/benefit} and \textit{social progress}, did not relate to the expected stakeholder desiderata. For instance, no pattern could be found indicating that the \textit{economic competition/development} frame was used more by funding stakeholders. A closer look at the specific occupations of participants that used this frame reveals it was most prominent for participants whose work focused on financial aspects within their stakeholder group. The participants who used this frame most were from different groups such as policy, media, and funding, but their occupations were all related to the financial side of these fields (E.g., a participant from the media and journalism group whose newspaper focused on finance-related news). The use of the other two frames of \textit{risk/benefit} and \textit{social progress} was distributed relatively evenly across all stakeholder groups. We observed no pattern that would indicate that these two frames are related to the ethics participants

The other frames also did not show notable overlap with specific stakeholder groups. The frames which make up \textit{comparison to other technology} were both equally spread out across stakeholder groups, as both \textit{comparison to traditional/classical technology} and \textit{comparison to AI} did not show a pattern towards certain stakeholder desiderata. The same holds for the frames of \textit{ones and zeros}, \textit{quantum computing},\textit{ quantum is complex/difficult}, and \textit{spooky/enigmatic}, although these were used relatively sparingly, so our claim is weak.

\subsubsection{Relation between future visions and stakeholder desiderata}
The use of hype and related concepts, such as alarmism, scepticism, nuance, use cases and future projections, did not show a pattern towards certain stakeholder groups nor a link with their desiderata. The same generally holds true for the different categories within use cases and future projections. Only for the media group were specific use cases more prevalent than non-specific ones. The content of the use cases does not show a pattern in relation to stakeholder desiderata, and only on a few occasions are these use cases related to the participant's occupation. The same observation holds for future projections. For instance, one participant who was asked about the effect of quantum developments on their line of work (related to ethics) said:
\begin{quote}
    [Ethics \#1] ``I think the field of quantum ethics is going to grow. I think it's going to increase, just out of sort of intuition, because that's what happened with AI ethics. So what that basically means is that I think the development of quantum technologies, I think it's probably going to give quantum ethicists more work."
\end{quote}
Another participant, when asked which quantum concepts would be important to their field, also related to their background while mentioning a milestone:
\begin{quote}
    [Funding \#3] ``So you have the quantum computer, which is hardware, but to be able to work with that, you would need those algorithms, which our research group is working on. That is the quantum software. Although I am a bit biased, I see it as very relevant. Because without such software, the hardware would never work."
\end{quote}

These occurrences do not seem to be linked to the stakeholders' desiderata, but rather to their individual backgrounds and occupations. Still, we do not observe a clear pattern between the use case and future projection framing and the participants' backgrounds. These future visions were generally present in discussions about quantum technology and its applications, but participants were not asked directly to give such examples. Instead, they brought them up of their own volition and as part of an explanation, or when asked more general questions about how they predicted the future of quantum technology.

\subsubsection{Relation between language use and stakeholder desiderata}
The use of metaphors is unevenly spread across the various stakeholder groups, with policy stakeholders using noticeably more metaphors than the other groups. Although these stakeholders use nearly every USAS semantic category (only omitting C, E, F, P and Z), the only categories they used significantly more than other groups were M (Movement, location, travel and transport) and X (Psychological actions, states and processes). Interestingly, only one metaphor this stakeholder group used could be related to category G (government and the public domain), in which an analogy was made between the day quantum encryption could be breached (Q-day) and D-day, which mentioned the German Empire. Overall, there seems to be little bias of specific metaphor categories towards certain stakeholder desiderata. 

Looking beyond metaphor \textit{categories}, it also seems that none of the stakeholder groups consistently focus on specific metaphors (E.g., none of the groups showed a proclivity toward using Schrödinger's cat). However, on occasion the specific background of stakeholders shines through in the use of metaphors. One participant, who worked for a newspaper and is accustomed to writing in a popularizing manner, explained quantum computer error correction in the following way:
\begin{quote}
    [Media \#1] ``I used a soccer team as comparison. In a regular ideal situation you would just send the normal amount of soccer players into the field, but with the qubit business you send out the entire first team including all the reserves and hope that maybe one will score a goal.``
\end{quote}
Another participant [Funding \#3], who had to work in two cities, described themselves as ``being in superposition``, and a participant [Startup \#1] who worked as a project manager described ``superposition like a project proposal in a draft phase``. These metaphors were never used more than once, and by more than one participant. A caveat to these findings is that only the source domains were coded and not the target concepts which these metaphors aimed to address. 

\textit{Use of jargon}
The use of jargon was unevenly spread across the various stakeholder groups, with both media and startups using relatively little jargon. Furthermore, the prevalence of technical knowledge was particularly high for the policy stakeholder group, relative to the other groups. The use of jargon seems related to the specific backgrounds of the participants, as for instance terms such as QKD and PQC were most prominent for a participant [Policy \#3] who worked closely with communications. A link can also be seen between knowledge of a subject and scepticism of its existence. For instance, one participant working with communications explained:
\begin{quote}
    [Policy \#3] ``Quantum internet does not exist. It does not exist. Quantum internet does not exist, but people would like it to exist."
\end{quote}

\section{Discussion}

In this study, we analysed the use of frames and language of non-expert QT stakeholders when explaining concepts from QST and discussing the future of this emergent technology. In doing so, we assessed the use of hype and stakeholders' desiderata to see how these concepts could shape frames and language use. Through a qualitative content analysis of 12 semi-structured interviews, we identified notable patterns that can form the basis of future studies.

\subsection{The language of non-expert QT stakeholders}
We observed that three important frames, previously identified by \citet{Meinsma2023}, dominated the conversations: \textit{risk/benefit}, \textit{social progress} and \textit{economic competition/development}. Within these three frames, we repeatedly saw two narratives previously identified by \citet{Suter2024} and \citet{Pohlmann2024}: ``Challenges to traditional cybersecurity" and ``National quantum strategies", respectively. Given that these previous studies also focused on communication by non-expert stakeholders, our findings imply that these two narratives might be a common theme among this stakeholder group. In the following discussion, any language that falls into either framing category of \textit{economic competition/development} or \textit{social progress} will be referred to as \textit{socio-economic} framing.

In our study, we found that non-expert stakeholders used both \textit{risk/benefit} framing and \textit{socio-economic} framing in almost equal frequency. Additionally, the participants in our study mostly showed a balanced view between the risks and benefits of QT. Regarding the potential benefits of QT, they showed reserved views. In some cases, they even expressed scepticism by explaining the \textit{socio-economic} context of QT development. Examples of these views are the idea that mostly the upper class would have access to the benefits of QT, or the idea that startups often make unrealistic promises to secure funding. 

This result contradicts the studied TEDx talks by \citet{Meinsma2023}, where an exploratory analysis showed that speakers in TEDx talks tend to emphasise the benefits of QT, with significantly less focus on both \textit{economic competition/development} and \textit{social progress}. The results also differ from an analysis of policy documents by \citet{Pohlmann2024}, which showed a prevalence for economic advantages over societal issues. However, the prevalence of the \textit{socio-economic} frame corresponds with the findings of \citet{Nisbet}, who identified that socio-economic framing was dominant in discourse around nuclear energy, evolution, and climate change. However, it is worth noting that \citet{Meinsma2023} explained the sparse occurrence of this frame by the strict coding used in their work, and they suggested that future inquiries might apply the frame more leniently. In this study, we applied the code less strictly.

Overall, the moderate use of the \textit{socio-economic} frame and the balanced view the participants held between the risks and benefits of QT seem to differ from previous research that investigated different forms of communication. The one-on-one interviews performed in this work show a new pattern compared to the popular communications investigated by \citet{Meinsma2023} and the policy documents analysed by \citet{Pohlmann2024}. 

One explanation for these differences is the setting in which QT was mentioned. Public speaking and engaging a large audience, as well as official documents stipulating possible policy, may stimulate very different language from one-on-one, private interviews. To hook a TEDx audience, a speaker might focus on the promising aspects of quantum and leave out more complex reflections on economic and societal dynamics, or even technical challenges. The goal of a TEDx speaker is to engage, while in our semi-structured interviews we gave space for in-depth discussions with follow-up questions. On top of that, we explicitly asked the stakeholders to imagine themselves in a day-to-day setting in their line of work, explaining QST to a colleague. It should be noted that the marginal nature of \textit{spooky/enigmatic} framing and \textit{mystical/holistic} framing we observed could also be explained with the above. 

\textit{Attention to quantum computing versus other quantum technologies.} In agreement with \citet{Meinsma2023}, we found that the QT stakeholders in this study mostly mentioned quantum computing in general discussions about QT. \citet{roberson2021talking} argued that responsible societal embedding requires discourse about multiple forms of QT, instead of a narrow focus on quantum computing and its applications.

However, the participants are aware of other forms of quantum technology and exhibited reasonable knowledge in these areas. Using this insight, the relatively frequent mention of quantum computing should not be interpreted as a lack of awareness of other forms. 

\subsubsection{Comparison to other technologies}

In this study, two new types of framing emerged from the data where participants contextualize QT through comparison to an existing (emergent) technology. This novel result emerged during our open coding process. Previous studies such as \citet{Wackers2025} and \citet{Meinsma2023} on communication around QT used quantitative, preset frameworks which excluded technical comparisons from data that might have contained this kind of framing. 

\textit{`Classical' technologies.} A recurring frame in almost every interview was the image of current technologies as ``classical" or ``traditional", as opposed to quantum technologies. Typically, this type of framing occurred when participants were asked to explain how different applications of quantum science worked, or how they might impact a future society. This ``classical" versus quantum technology rhetoric implies that participants imagine a future reality where QT will be the new ``normal", and where our current stage of technological advancement would belong to a previous technological age. 

It is possible that the term ``classical" is related to the distinction between classical physics and quantum physics: \citet{Vermaas2017} described how in the early 20th century, quantum physicists started to label 19th century physics as ``classical" in order to highlight the counter-intuitive nature of the newly discovered quantum mechanics. It is then argued that these contrasting notions between ``classical" physics and the counter-intuitive, enigmatic quantum physics still exist today, which raises the barrier for stakeholders to discuss QT. We recommend future studies to analyse how the ``classical"-distinction from physics found its way into non-expert communication in QST, and how the rhetoric of ``classical" technology might contribute to discourse around QT.

\textit{Artificial Intelligence as a benchmark.} The mention of/comparison to AI was one of the most dominant frames in our data. In some cases, this was in a technical context. We saw a recurring notion of AI and QT enhancing each other, which is a relevant topic within scientific literature \citep{Alexeev2025-md, Devadas2025}. However, we mostly saw that participants used the revolutionary impact of AI as a ``benchmark" to compare any possible future impact of QT to. The reason participants often mentioned AI might be that, similar to AI, quantum technology is also an emerging technology that has been widely reported on and is expected to have a large global impact \citep{Cerutti2025, NaturePortfolio2025}. In other words, this finding suggests that stakeholders use AI to make sense of the potential implications of QT.

\textit{Continual approach to (emerging) technologies?} The significance of the above results is that their occurrence was not prompted by the interviewers, as we asked questions about QT, not AI. To our knowledge, existing literature on framing of QT rarely touches on the connection between QT and AI. One of the studies that does make a connection between QT and AI is \citet{godoy-descazeaux_images_2023}. However, their connection is less about understanding and making sense of the potential implications of QT and more about illustrating how quantum computers can speed up the development of AI.

While not explicitly making a connection between QT and AI, other studies suggest that reference to other technologies is made in pursuit of understanding QT \citep{Wackers2025, lockwood_giving_2022}. In this light, the connection between QT and AI can be interpreted as the stakeholders' effort to use a familiar concept to understand the potential implications of QT. This makes sense because our non-expert stakeholders are likely to be exposed to AI through their occupations.

This effort to understand the implications of QT through reference to AI seems to be in line with \citet{Shelley-Egan2025-oi} that argues for the continuity approach in the analysis of QT's potential impact. In the context of ethics discourse around QT, \citet{Shelley-Egan2025-oi} argue that ethical concerns around QT are mistakenly treated as an entirely novel occurrence, which creates ``speculation and misplaced resources and energy" (p.1). Building from this idea, they promote viewing QT through a lens of continuity within a broader idea of new and emerging science and technology (NEST). While their study covers a broader ethical discourse around QT, we believe the idea can also be applied to analyse communicative elements related to understanding explanation. Many of the participants in our sample are not solely occupied with ``quantum", but also with the overall impacts of emerging technologies such as AI. Therefore, we advise including framing that compares QT with AI or another emerging technology in future studies of communication around QT. Consequently, QT is acknowledged as one part of the continuous collection of NEST.

\subsection{Hype, alarmism and future visions}
\textit{Individual occurrences of hype and alarmism.}
This research found that direct instances of hype and alarmism when discussing QT were less prevalent than would be expected from literature \citep{Meinsma2023, WangXu2026_typeshype, Roberson2023, Meyer2023}. One possible reason for this stems from the setting in which participants would normally talk about QST. As this discourse exists in more colloquial occupation-related channels, where QST is discussed between coworkers as part of a daily routine, the approach to this topic and the audience differ from those in previously investigated TEDx talks or policy papers. Furthermore, as discussed in the previous section on frames, this difference might also stem from the research setting, as hype might be more prominent in a speech or in casual conversation than in an interview, whilst nuance could be more prominent in prolonged interviews. Current literature does not seem to have investigated this phenomenon, which could be a possible avenue for further research. Although the participants were asked to speak with us as if we were colleagues, they might still converse differently from how they actually would with their colleagues, and the interviews might have led participants to be more cautious about hyped expressions because of a social desirability bias \citep{Atkeson2014}.

Although the use of alarmism was more pronounced than hype, the difference is too small for the limited qualitative research to draw conclusions from. As current research into the prevalence and context of alarmism is limited, future research might investigate how hype and alarmism are expressed and linked. As alarmism had a significant overlap with the \textit{risk/benefit frame} and this narrative is also found to be prevalent in society \citep{Intemann2022-en}, it might be interesting to pursue further inquiries into the prevalence of alarmism. This could be especially promising as one participant [Startup \#2] focused on the risks of government control through the use of QT, echoing similar worries raised during the COVID-19 pandemic \citep{Hrbkov2024}.

\textit{Use cases and future projections.}
The framework for coding hype through use cases and future projections, as envisioned by \cite{karami_quantum_2026}, was applicable to a wide variety of instances. Although physical applications were most prominent, which is what the framework was originally designed for, the fact that participants also brought up social issues shows that this framework is versatile in the topics it can be used for. This indicates that hype has reached beyond technical applications to wider societal implications \citep{roberson2021talking}. Both use cases and future projections can be used to code not only for hype, but also alarmism or even nuance, while allowing for a wide field of both social and technological topics. This versatility makes it excellent for future inquiries, although it must be noted that a more precise definition of the difference between a ``specific", ``rather specific" and ``non-specific" use case is warranted if this framework were to see widespread application. Furthermore, although most of the use cases that were brought up corresponded to common ones found in the literature \citep{Ukpabi2023, Singh2022}, an expanded framework could also include examples of non-technical use cases .

There are a multitude of ways to understand how framing and language are used in combination with use cases and future projections. The significant overlap between use cases and the frames \textit{comparison to other technologies}, \textit{risk/benefit} and \textit{social progress} is expected, as these were used as tools to explain possible use cases, concurring with literature \citep{SchwarzPlaschg2018, Jasanoff2015}. Although these use cases were mostly positive, the fact that risks and nuance came up shows that this positive nature is not necessarily synonymous with hype. Although we would expect non-specific use cases that could lead to hype, such as vague yet overtly positive applications of technology, this was not the case, as our findings suggest that even the non-specific use cases were often presented in a nuanced way.

One interesting pattern that we observe is between specific use cases and jargon. One possible explanation is that stakeholders with moderate to high technical knowledge of specific quantum topics were able to give more precise examples of their applications. In turn, their knowledge also allowed them to recognise related jargon, which is relevant or even required to describe a specific use case. These specific use cases and the related language use did not seem to relate to hype, possibly indicating that people with high technical knowledge are less prone to hype in the setting of these interviews. However, that is not to say that people with a high degree do not fall prone to hype, which has been observed in the past \citep{Intemann2022-en}. 

In contrast, the overlap between the non-specific use cases and the \textit{because of quantum} frame could both imply a lack of (technical) knowledge on the side of the participant. Furthermore, the fact that participants also try to come up with use cases based on a name (A quantum computer or quantum sensor having the same use cases as a classical computer or sensor), might indicate how people venture a guess when they are unsure, something also found in literature \citep{Bertola2025}. However, as we show in the previous section, this overlap is observed when participants lack knowledge of possible applications of QT. Interestingly, they offer a nuanced disclaimer that they do not know enough about the subject. It thus seems that it is not as straightforward to say that a lack of knowledge would lead to more hype, since a nuanced answer is also observed. 

\subsubsection{Unexpected future visions: Socio-economic nuance and scepticism}

We did not anticipate the significant presence of a nuanced approach to future predictions regarding QT, or the appearance of scepticism. To our knowledge, the literature rarely touches on this nuanced framing of QT, and when it does, the nuanced view is the minority. For example,  \citet{Meinsma2023} reported that only 4\% of TEDx talks on QST approach the topic with a balanced view of the benefits and risks of QT. Our findings seem to suggest the contrary.

In the literature, a nuanced view is often indicated by discussion of both benefits and risks \citep{oehmer-pedrazzi_content_2023}. Our study enriches this discussion by identifying two modes of voicing, offering a nuanced view. The first is centred on technical limitations, which leads to a constrained view of technical risks and benefits of QT. The second, which is more prominent in our data, is approached through \textit{socio-economic} framing. Additionally, scepticism generally occurred through the frame of \textit{economic competition/development} (examples in Section \ref{results_scepticism}). 

However, when looking at the use cases mentioned in the interviews, non-specific use cases of QST dominated over rather specific or specific use cases. Given that participants were generally hesitant to endorse highly positive notions about the future of QT, this paints an interesting picture of how these stakeholders deal with hype. In the framework of \citet{karami_quantum_2026}, unspecified future visions tend to coincide with hype because, in their dataset of scientific literature, scientists resorted to vague claims presumably to attract the attention of other stakeholders. However, our findings seem to warrant more investigation into the framework. In our study, unspecified claims about QT cannot be interpreted merely as hyping, in the sense of gathering resources and support. This result warrants a review of how the literature connects QT stakeholders' future visions to the concept of hype. 

Moreover, the manner in which non-expert stakeholders approach discourse around QT touches on a discussion around understanding of QT. \citet{De_Jong2026-vg} argued that not a \textit{technical} understanding, but a \textit{functional} understanding is necessary to engage in the broader socio-economic debate around QT. Functional understanding encompasses insight into ``what it can do and what it can be used for". A grasp of the technical aspects of QT can provide some insight into ethical concerns, but is not sufficient or needed to understand how the technology will interact with society. In this study, we show that non-expert stakeholders are well capable of engaging in a discussion of the ethical and societal aspects of QT, while their ideas of use cases are not highly detailed - which would require a deeper technical understanding. Thus, this study supports the ideas of \citet{De_Jong2026-vg}.

The tendency towards socio-economic nuance and the resistance against engaging with hype complement previous studies. For example, 24 quantum researchers interviewed by \citet{SotoSanfiel2025} pointed to the media, major corporations, and scientists themselves as contributors of hype. They suggested that hype occurs due to systemic pressures within the scientific system, which is also in line with \citet{caulfield_science_2012}. Interestingly, \citet{SotoSanfiel2025} indicated that scientists might use hyped language to convince policymakers, even though the government participants we interviewed take a nuanced approach in their communication, suggesting they might not be prone to hype. This suggests a difference in the use of hype within and across stakeholder groups, coinciding with research on differences between intra-group and inter-group communication styles \citep{Communication_Accommodation_Theory_2015}. Our study suggests that nuance is more prevalent in in-group communication, indicating that these non-expert stakeholders are aware of exaggeration and hype in the field. Future research might investigate how hype occurs even when the non-expert audience exhibits nuance and caution in evaluating exaggerated claims.

There are multiple arguments as to why a reserved stance on QT might have been prevalent in this sample of stakeholders. First, the method might have stimulated a more nuanced discussion: as described previously, a one-on-one interview of approximately one hour creates space for participants to engage in nuanced discussion. With an audience of just one interviewer and a private setting, participants might not feel the need to ``convince" their audience of the benefits or potential of QT.

Another related potential cause of the observed nuanced views is that most participants' occupations typically do not require them to engage in hype to foster engagement and investments. In the interviews, we explicitly instructed the participants to imagine themselves talking to a colleague on a typical workday, which could have further weakened any incentive to engage in hyped discourse around QT. Their occupation requires them to, for example, advise the government, inform the general public, or reflect on the ethical side of QT development . On the other hand, it is reasonable to expect that participants working at a startup might show relatively greater hype. However, one participant in this group engaged in mostly alarmism, scepticism and only some nuance - no instances of hype were coded. Our other participant in this group showed mostly nuance, and only one instance of hype. Therefore, the link to stakeholders' occupation may not be clear and needs further study. 

\subsection{Effect of stakeholder desiderata}
\subsubsection{\textit{The link between stakeholders and their desiderata}}
The results indicate that it is not straightforward to link stakeholders to their desiderata through their language use, contrary to what would be expected from \citet{Umbrello2024}. Here, a distinction can be made between the motivation and description stakeholders gave of their occupation, which generally overlapped with the desiderata, and the frames and language use. The only contradiction between the self-description and the expected desiderata was evident in one startup participant who did not care for the completion of quantum computers, instead focusing solely on their own livelihood. What is important to note is that the desiderata for a stakeholder group (Table \ref{tab:stakeholder desiderata table}) are written in such general terms as to reflect their reasoning and considerations. For example, an ethics expert in QST would be unlikely to disregard the importance of ethics in QST. Thus, the use of these desiderata is limited in understanding why stakeholders use specific language and frames, as people are unlikely to deviate from their expected desiderata due to conformity bias, especially when interviewed by outsiders \citep{Mneimneh2012InterviewPA}.

Relating the desiderata to frames and language revealed shortcomings in \citet{Umbrello2024}'s approach. Although it would be expected from the desiderata that the startups had the least concern for the frames of \textit{social progress} and \textit{risk/benefits}, while those in ethics would show the most interest, the fact that this frame was spread relatively evenly across all participants supports the idea that all stakeholder groups were at least aware of the social or ethical impact of QST and the importance of facilitating this. A possible explanation for this result might be conformity bias, as people might not admit in a 1-on-1 interview that they do not care about the risks posed by the technology they are developing \citep{Padalia2014}. Furthermore, there is often a difference between stakeholders' concerns and their actions, the so-called intention-behaviour gap \citep{Hassan2014}.  Of course, it may also be that most stakeholders simply care about the ethical implications.

Overall, the small size of the startup stakeholder group does not allow for statements about their ethical biases, especially given that this study focused mostly on employees, while there is a large difference between the ethical attitudes of employees and employers \citep{Kim2022}. The stakeholder groups of \cite{Umbrello2024} do not allow for a distinction between various levels of occupation within a group; differing desiderata within a stakeholder group might also explain why the link between desiderata and frames/language use is difficult to make.

An interesting pattern we observe is media stakeholders' use of use case framing. This might be understood by the fact that media needs to provide concrete examples to the public, which may be less relevant for other stakeholder groups.

\subsubsection{\textit{The link between stakeholders and their occupation}}
A more convenient way to observe possible biases in language and framing might be through the stakeholders' occupations. For instance, stakeholders in policy and media would not be expected to bring up finances significantly more than other groups. However, the fact that this occurs for specific stakeholders is understandable, considering that their occupations are at a financial newspaper. In a similar vein, the distance a startup stakeholder had to the finalisation of a quantum computer can be understood through the fact that their occupation (and livelihood) depends more significantly on the development of such a machine.  For use cases and future projections, the overlap was also closely related to the stakeholders' occupation.

Although the prevalence of metaphors for certain stakeholder groups does not seem to be explained by either their desiderata or occupations, this might simply be an artifact of the small sample size. However, a closer analysis that splits the metaphors into more general ones (Schrödinger's cat, a coin spinning) and specific ones (a soccer team, a proposal in a draft phase) suggests a connection between an occupation and its preferred metaphors. Such a link between occupation and metaphors is also present in the literature \citep{Woodhams2014}. Furthermore, certain instances of jargon, such as QKD and PQC, could be directly linked to stakeholders' occupations. It thus seems that occupation is a better way of understanding a stakeholder's frames and language use than their desiderata.

Another notable issue is that stakeholder groups did not always clearly overlap with their stakeholder occupations. As previously mentioned, certain participants in policy and media might be more related to finances, whereas funding participants saw themselves mostly as communicators and connectors, which wasn't covered by any of the stakeholder groups. Within the startup group, different occupations, such as project managers or technicians, also have different interests. This shows a juxtaposition between the theoretical basis of the envisioned stakeholder groups and their actual application. 

\subsubsection{\textit{How stakeholders form knowledge, and apply this knowledge in frames} }
Instead of looking at desiderata or occupation to see how jargon is formed, it is also possible to look more specifically at how people gathered knowledge. Most of the participants mentioned that they obtained their knowledge of QST through their occupation, via colleagues or recommendations for what to read/watch. Furthermore, although most of the participants did not have knowledge of QST through their education, they were largely highly educated, which is linked to a pursuit of further knowledge \citep{Boeren2017}. This might also explain the prevalence of technical knowledge among these non-experts stakeholders. As such, it seems that stakeholders' occupations form the basis for knowledge development. This is further supported by the fact that the technical knowledge that certain participants had on QT was directly linked to their occupation. Furthermore, the knowledge people obtain in their occupation is not limited to technical knowledge and jargon; it also encompasses social interactions and can take a culturally shaped form, which might create a bias towards the dominance of certain frames used in workplaces \citet{Tynjl2008, AChatman2022}.

Overall, it seems that simply using the desiderata to understand framing and language use falls short of a comprehensive explanation. Instead, it is important to consider a combination of  stakeholders' backgrounds, work environments, and occupations. Only by considering these factors collectively can we gain a deeper understanding of the framing and language used by non-expert stakeholders.


\section{Conclusion}
The increase in interest in QT in recent years has warranted the involvement of stakeholders, and an evaluation of their position in the quantum ecosystem. To facilitate stakeholder understanding, this research set out to investigate how the prevalence of frames, language use and hype by non-expert quantum stakeholders could be understood. To answer this query, 12 interviews with non-expert quantum stakeholders from 5 stakeholder groups were held. 

The interviews revealed the prevalence of existing and new frames, metaphors, and jargon relating to the discussion of QST. The equal use of the \textit{socio-economic} and \textit{risk/benefit} frames contradicts previous studies \citep{Meinsma2023, Pohlmann2024} in which the \textit{socio-economic} frame was much less prevalent. These seemingly contradictory findings might be explained by methodological differences, as one-on-one interviews give participants more space. Analytically, this difference can be understood in light of the stakeholders' backgrounds and occupations. 

Furthermore, a comparison of QT with ``classical" technology and AI was also found to be prevalent. This fits the broader discussion suggesting that an existing technology is used to understand the impact of an emerging technology \citep{Wackers2025,lockwood_giving_2022}. This finding suggests that further research can use the continuity lens to analyse the communication of QT \citep{Shelley-Egan2025-oi}, which positions QT as an emerging technology among many that come and go.

We also found that the prevalence of hype and alarmism was much smaller than that of nuance. In contrast to previous studies \citep{Meinsma2023, WangXu2026_typeshype, Roberson2023, Meyer2023}, Nuance is the unexpected frame that we found in our study. This suggests that our non-expert stakeholders might not be prone to hype. While this finding can be attributed to our method of choice, interviews, it seems to be in line with previous research that is due to systemic pressure \citep{caulfield_science_2012}, not personal motive. As these non-expert stakeholders work in a hyped field, this finding calls for further research into nuance, as they seems prone to hyping practices, despite

Lastly, a framework by \cite{Umbrello2024} was tested relating to stakeholder desiderata. Although this framework is useful for identifying and characterising quantum stakeholders, it failed when we tried to relate the expected stakeholders' desiderata to the language they used and the future visions they held. We found that focusing on stakeholders' occupations yields a more fruitful understanding of their framing and language use.

This study is intended as a stepping stone for future research, not only enabling concepts such as hype and framing to be better analysed, but also facilitating further investigation into the interests of quantum stakeholders and society as a whole.

\subsection{Limitations and future research} 
Certain developments during this research are important to consider when reading these results and conclusions. The largest issues pertain to the stakeholders that were contacted and interviewed. Only one stakeholder was not affiliated with a Dutch organisation, and many were connected to QDNL. Quantum stakeholders at the national security agencies are difficult to reach, as were corporate institutions such as banks and larger companies. This might influence the results, and future research should look to contact stakeholders through more varied sources. 

Furthermore, the “general public” stakeholder group was also omitted, minorities were under-represented, and only one stakeholder did not hold a university degree. Although from a research perspective this does not hinder interesting results from forming, it is lacking in considerations of diversity, equity, and inclusion, which should be accounted for to fairly include all stakeholders and ensure research benefits all of society \citep{Vermaas2017, ColnAguirre2022, Luyendijk2022-op}

It must also be acknowledged that our method of choice, interviews, might shape our findings.
The differences between a speech to a crowd versus a one-on-one interview likely caused the participants to discuss topics with more nuance than hype. Instead of giving a quick, positive, and hyped answer, the participants could discuss topics in greater depth and, as a result, bring a more nuanced view. Furthermore, although this research aimed to emulate the natural ways in which quantum stakeholders might converse with each other, we had no measure to confirm this. The interview setting possibly still caused the participants to behave differently than they normally would. Future research could explore different methods to understand how stakeholders communicate with each other in their natural environment.

Despite its limitations, this research identified a variety of avenues for future research. Both the rhetoric of ``classical" technology and comparisons to AI can be relevant when investigating frames raised in discussions surrounding QST.  Furthermore, the prevalence of frames related to hype, such as alarmism, scepticism and nuance, warrants further scrutiny as these are largely under-represented in literature yet emerged as important factors in these interviews. Another noteworthy finding is that members of certain groups, such as scientists and policymakers, are less likely to use hype within their group, while they do recognise the function of hype when communicating to the other group. Future research could investigate how stakeholders use hype differently between in-groups and out-groups. Lastly, an overview of quantum stakeholders that is better aligned with their occupations rather than their stakeholder desiderata could shed light on the interests different stakeholders have in QST.

%
%


\newpage 
\ack{
Floris Löffler and Lisanne van Veenen would like to acknowledge and thank their supervisors and fellow students. This research, and the pleasant work environment that enabled it, would not have been possible without the positive and accepting attitudes of Julia Cramer and Muhammad Unggul Karami. Their aim to not only aid us in our research, but also to ensure we did not overextend ourselves, greatly contributed to the work we performed and the regard in which we felt we were held. We would also like to thank our fellow students, who not only aided us with advise in our similar ordeals, but also made our stay at the Q\&S group much more fun. We would also like to thank all of the other members of the Q\&S group, for including us in the world of science communication and as one of their own. Lastly, the first two authors would like to acknowledge each others work, contributions and cooperation during this project.
}

\funding{
This work was partly supported by the Dutch National Growth Fund (NGF), as part of the QDNL programme.
}

\roles{
\begin{itemize}
    \item FL: Conceptualization, Data Curation, Formal Analysis, Investigation, Methodology, Project Administration, Writing – original draft, Writing – review \& editing
    \item LvV: Conceptualization, Data Curation, Formal Analysis, Investigation, Methodology, Project Administration, Writing – original draft, Writing – review \& editing
    \item MUK: Conceptualization, Supervision, Writing – review \& editing
    \item JC: Conceptualization, Funding Acquisition, Supervision, Writing – review \& editing
\end{itemize}
The first two authors performed the research and wrote the report in equal measure.
}

\data{
All the data gathered from the interviews is accessible in anonymised form upon contact via the correspondence address. For more data and forms relevant to this project, see the SI.
}

\bibliographystyle{agsm} 
\bibliography{references}

\end{document}